\documentclass[12pt]{iopart}
\usepackage{graphicx}
\usepackage{iopams}

\begin{document}

\title{Radiative energy loss reduction in an absorptive plasma}

\author{M Bluhm$^{1}$, P B Gossiaux$^{1}$ and J Aichelin$^1$}

\address{$^1$ SUBATECH, UMR 6457, Universit\'{e} de Nantes, 
Ecole des Mines de Nantes, IN2P3/CNRS. 4 rue Alfred Kastler, 
44307 Nantes cedex 3, France}

\ead{bluhm@subatech.in2p3.fr}

\begin{abstract}
The influence of the damping of radiation on the radiative energy 
loss spectrum of a relativistic charge in an infinite, absorptive 
plasma is studied. We find increasing reduction of the spectrum 
with increasing damping. Our studies, which represent an Abelian 
approximation for the colour charge dynamics in the quark-gluon plasma, 
may influence the analysis of jet quenching phenomena 
observed in high-energy nuclear collisions. Here, we focus on 
a formal discussion of the limiting behaviour with increasing radiation 
frequency. In an absorptive (and polarizable) medium, this is determined 
by the behaviour of the exponential damping factor entering the spectrum 
and the formation time of radiation. 
\end{abstract}


\section{Introduction \label{sec:1}}

The strong quenching of jets observed experimentally in high-energy 
nuclear collisions at the Relativistic Heavy Ion 
Collider~\cite{Arsene05,Back05,Adams05,Adcox05} 
and the Large Hadron Collider~\cite{Aad10} 
has been interpreted as an indication for the 
formation of an opaque, deconfined plasma of QCD matter, in which the 
leading partons suffer medium-induced energy loss. The energy loss 
of relativistic partons is expected to be dominated by gluon 
radiation~\cite{Gyulassy94,Baier97,Baier00}. Important in this context is the possibility 
of a destructive interference between the radiation amplitudes of 
multiple deflections during the formation time of gluons~\cite{Baier95} 
and the influence of the dielectric polarization of the 
plasma~\cite{Kampfer00,Djordjevic03}. 
Both effects lead to a modification of the gluon radiation spectrum and, 
consequently, of the radiative energy loss. So far, however, the possible 
influence of damping of radiation in an absorptive plasma has been 
neglected in these considerations. 

\section{Radiative energy loss spectrum \label{sec:2}}

Recently~\cite{Bluhm11}, the impact of polarization and absorption effects 
in a medium on the radiative energy loss of a relativistic 
charge $q$ has been studied in linear response theory. Thereby, the 
influence of multiple deflections of the charge is incorporated in line 
with the original approach in~\cite{Landau53}. These investigations can be 
viewed as abelian approximation for the dynamics of a colour charge in the QCD 
plasma. Nonetheless, they miss still the important non-abelian contributions 
from gluon rescatterings~\cite{Baier95}. The absorptive, dielectric medium is 
phenomenologically modelled by a complex index of refraction squared of the form 
$n^2(\omega)=1-m^2/\omega^2 + 2i\Gamma/\omega=[n_r(\omega)+in_i(\omega)]^2$ with 
frequency $\omega$, and $m$ and $\Gamma$ accounting for a finite in-medium gluon 
mass~\cite{Kampfer00,Djordjevic03} and damping 
rate~\cite{Pisarski89,Peshier04}, respectively. 
For an infinite medium with permeability $\mu(\omega)=1$, and in the case $n_r$ 
and $n_i$ have equal signs, the radiative energy loss spectrum per 
unit length reads~\cite{Bluhm11} 
\begin{equation}
\label{equ:1}
 -\frac{d^2W}{dzd\omega} \simeq -\frac{2\alpha}{3\pi} \frac{\hat{q}}{E^2} 
 \int_0^\infty d\bar{t} \,\omega \cos(\omega\bar{t}\,) 
 \sin\left[\omega|n_r|\beta\bar{t}\left(1-\frac{\hat{q}\bar{t}}{6E^2}\right)\right] 
 \mathcal{F}(\bar{t}\,) \,.
\end{equation}
Here, with $\hbar=c=1$, $\alpha=q^2/(4\pi)$ is the coupling, 
$\beta=\sqrt{1-1/\gamma^2}$ with $\gamma=E/M$ for a charge with energy $E$ and 
mass $M$, 
$\mathcal{F}(t)=\exp[-\omega|n_i|\beta t(1-\hat{q} t/(6E^2))]$ 
is the exponential damping factor related to the imaginary part of $n(\omega)$ 
and the parameter $\hat{q}$ denotes the mean accumulated transverse momentum 
squared of the deflected charge per unit time. 

\begin{figure}[t]
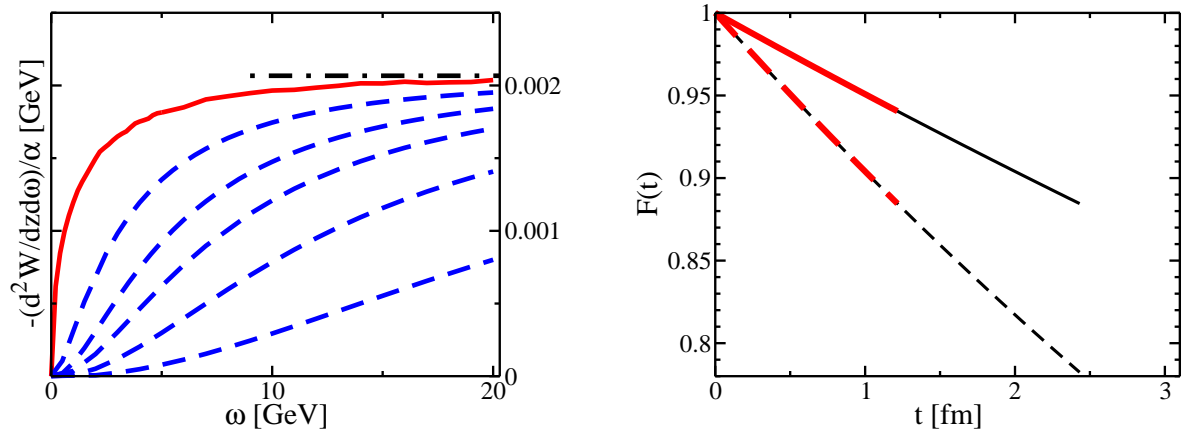

\vspace{7mm}
\begin{centering}
\includegraphics[scale=0.32]{spectrum14_1.eps}
\hspace{7mm}
\includegraphics[scale=0.32]{ExpFactor3.eps}
\caption{\label{Fig:1} Left: radiative energy loss spectrum 
per unit length as a function of $\omega$ for a charge with $E=50$~GeV, 
$M=4.5$~GeV and $\hat{q}=1$~GeV$^2/fm$. The solid curve depicts the vacuum 
result Eq.~(\ref{equ:2}) as a reference, while the dash-dotted curve shows 
its formal limit for increasing $\omega$, Eq.~(\ref{equ:add}). 
The dashed curves exhibit the 
full, reduced spectrum Eq.~(\ref{equ:1}) in an absorptive medium with $m=0$ 
and $\Gamma=10,\,20,\,30,\,50,\,100$~MeV from top to bottom. Right: exponential 
damping factor $\mathcal{F}(t)$ as a function of $t$ within the essential 
time interval of the integral in Eq.~(\ref{equ:3}) determined by $t_f$. 
The kinematic parameters $E$, $M$ and $\hat{q}$ 
are chosen as in the left panel. Upper solid curves are for $\Gamma=10$~MeV and 
lower dashed curves for $\Gamma=20$~MeV, while the thinner curves are for 
$\omega=20$~GeV and the thicker curves on top of them are for $\omega=40$~GeV.}
\end{centering}
\end{figure}

In the vacuum limit, i.e., when setting $n_r=1$ and $n_i=0$, the dominant 
contribution of Eq.~(\ref{equ:1}) for $\beta$ close to $1$ reads 
\begin{equation}
\label{equ:2}
 -\frac{d^2W}{dzd\omega} \simeq \frac{2\alpha}{3\pi} \frac{\hat{q}}{M^2} 
 \int_0^\infty dx \sin\left[x + \frac{2\hat{q}x^2}{3\omega M^2 
 (1-\beta^2)}\right] .
\end{equation}
This coincides with the result for the radiation intensity derived 
in~\cite{Landau53}, when $\hat{q}$ is identified with the parameters used 
in~\cite{Landau53}. 
With increasing $\omega$, Eq.~(\ref{equ:2}) approaches, for $\hat{q}E/M^4\ll 1$, 
the $\omega$-independent limit 
\begin{equation}
\label{equ:add}
 -\frac{d^2W}{dzd\omega} \simeq \frac{2\alpha\hat{q}}{3\pi M^2} \,. 
\end{equation}
The limiting behaviour Eq.~(\ref{equ:add}) 
as well as the full vacuum spectrum Eq.~(\ref{equ:2}) are 
shown for specific kinematic parameters in Fig.~\ref{Fig:1} (left panel) by 
dash-dotted and solid curves, respectively. 
As evident, the limit Eq.~(\ref{equ:add}) is approached rather quickly by 
the full vacuum spectrum. 

In Fig.~\ref{Fig:1} (left panel), also the full spectrum Eq.~(\ref{equ:1}) 
for an absorptive medium with $m=0$ and different, finite values of $\Gamma$ 
is exhibited by dashed curves. Even for small $\Gamma$, the spectrum is 
significantly reduced compared to the vacuum result in the region of small and 
intermediate $\omega$. This suppression of the spectrum is a consequence of 
the exponential damping factor in Eq.~(\ref{equ:1}), cf.~the 
discussion in~\cite{Bluhm11}. In 
the limit $\omega\gg\Gamma$ and $m=0$, the dominant contribution to 
Eq.~(\ref{equ:1}) for $\beta$ close to $1$ is given by 
\begin{equation}
\label{equ:3}
 -\frac{d^2W}{dzd\omega} \simeq -\frac{\alpha}{3\pi} \frac{\hat{q}}{E^2} 
 \int_0^\infty d\bar{t} \,\omega 
 \sin\left[\omega\bar{t}(\beta-1)-
 \frac{\omega\beta\hat{q}\bar{t}^{\,2}}{6E^2}\right] 
 \mathcal{F}(\bar{t}\,) \,. 
\end{equation}

It is interesting to study whether this damped spectrum 
Eq.~(\ref{equ:3}) formally approaches the same limit Eq.~(\ref{equ:add}) 
for increasing $\omega$ as the vacuum result. For 
this to happen, $\mathcal{F}(t)$ would have to become $1$ with increasing $\omega$ 
within the time interval, which is essential for the integral in 
Eq.~(\ref{equ:3}). In the considered limit $\omega\gg\Gamma$, $\omega|n_i|\to\Gamma$, such 
that $\mathcal{F}(t)\to\exp[-\Gamma\beta t(1-\hat{q} t/(6E^2))]$, which is 
{\it per se} $\omega$-independent as long as $\Gamma$ does not depend on 
$\omega$ and, thus, would imply a modification of the spectrum for any $\omega$. 
However, the essential time interval itself depends on the frequency, such that 
the influence of $\mathcal{F}(t)$ on the spectrum Eq.~(\ref{equ:3}) varies with 
$\omega$. The essential time interval may be determined by the 
formation time $t_f$ of radiation, which for not too large $\Gamma$ 
and/or $\gamma$, 
cf.~the discussion in~\cite{Bluhm11,Bluhm11tf}, can be found from 
Eq.~(\ref{equ:3}) by a condition for the phase factor 
\begin{equation}
\label{equ:4}
 \Phi(t_f)\equiv\omega t_f(1-\beta) + \frac{\omega\beta\hat{q}}{6E^2}t_f^2 \sim 1 \,.
\end{equation}
A rough estimate for the solution $t_f$ of this condition equation is given by 
the minimum of $t_f^{(s)}\sim 1/[\omega(1-\beta)]$ and 
$t_f^{(m)}\sim E\sqrt{6/(\omega\beta\hat{q})}$. For $\gamma$ 
small compared to $M^3/\hat{q}$, this minimum is, with increasing $\omega$, given by 
$t_f\sim t_f^{(s)}$. Then, with $\mathcal{F}(t)\sim\exp[-\Gamma t]$ for 
relativistic particles, the exponential damping factor decreases monotonically from 
$1$ at $t=0$ to $\exp[-\Gamma/(\omega(1-\beta))]$ at $t\sim t_f$. 
The behaviour of $\mathcal{F}(t)$ as a function of 
$t$ is exhibited in Fig.~\ref{Fig:1} (right panel) for different $\omega$ and different 
$\Gamma$. 
As evident, increasing $\Gamma$ will increase the influence of 
$\mathcal{F}(t)$ on Eq.~(\ref{equ:3}). With increasing $\omega$, $t_f$ decreases and, 
consequently, the decrease of 
$\mathcal{F}(t)$ plays less and less role in Eq.~(\ref{equ:3}). For $\omega\to E$, 
$\mathcal{F}(t_f)\to\exp[-\Gamma E/M^2]$, which is in general different from $1$. 
This implies that the asymptotic limit Eq.~(\ref{equ:add}) can be recovered for large 
$\omega$ only provided $E\ll {\rm min}\left(M^4/\hat{q}, M^2/\Gamma \right)$. 

For $\gamma\gtrsim M^3/\hat{q}$, 
$t_f\sim t_f^{(m)}$ even for $\omega\to E$, such that 
$\mathcal{F}(t_f)\sim\exp[-\Gamma E/\sqrt{\omega\hat{q}}]\to
\exp[-\Gamma\sqrt{E/\hat{q}}]$ for $\omega\to E$, implying qualitatively the same 
behaviour of $\mathcal{F}(t)$ as discussed before. Moreover, this illustrates 
the increasing importance of the exponential damping factor on the spectrum with 
increasing $E$, cf.~\cite{Bluhm11}. 

Similar observations can be made in the case of a medium with $m\ne 0$ 
in the limit $\omega\gg m\gg\Gamma$. Then, $n_r(\omega)\to 1$ rapidly with 
increasing $\omega$ such that Eq.~(\ref{equ:1}) reduces likewise to 
Eq.~(\ref{equ:3}) for $\beta\to 1$. Details in the limiting behaviour depend, 
though, on details in the formation time, which is influenced 
by finite $m$-effects, cf.~\cite{Bluhm11tf}. 

\section{Conclusions, and Acknowledgments \label{sec:4}}

In summary, we have presented results for the radiative energy loss 
spectrum of a relativistic charge undergoing multiple scatterings 
in an absorptive medium. Our investigations may be viewed as abelian 
approximation for the dynamics of a colour charge in an infinite 
deconfined QCD plasma. But they are applicable also, if the system 
size is large compared to the formation length of radiation. These 
studies, however, still miss the important non-abelian contributions 
from gluon rescatterings. We find an increasing reduction of the 
spectrum compared to the vacuum case for increasing damping of 
radiation in a purely absorptive plasma. In this paper, special 
emphasis is put on the discussion of the limiting behaviour of the 
spectrum with increasing $\omega$. For an absorptive medium, this is 
driven by the behaviour of an exponential damping factor in the 
spectrum, which is related to the formation time of radiation in 
the matter. 

We gratefully acknowledge valuable and insightful discussions with 
B.~K\"ampfer, T.~Gousset and S.~Peign\'{e}. The work is supported by 
the European Network I3-HP2 Toric and the ANR research program 
``Hadrons@LHC'' (grant ANR-08-BLAN-0093-02).

\end{document}